\documentclass[fleqn,twoside]{article}
\usepackage{espcrc2}

\makeatletter
\def\simleq{\mathrel{\mathpalette\gl@align<}}
\def\simgeq{\mathrel{\mathpalette\gl@align>}}
\def\gl@align#1#2{\lower.6ex\vbox{\baselineskip\z@skip\lineskip\z@
     \ialign{$\m@th#1\hfill##\hfil$\crcr#2\crcr\sim\crcr}}}
\makeatother

\newcommand{\bra}{\langle}
\newcommand{\ket}{\rangle}
\newcommand{\braket}[1]{\bra #1 \ket}
\newcommand{\qq}{\braket{\bar{q}q}}
\newcommand{\sbs}{\braket{\bar{s}s}}

\newcommand{\qGq}{g\braket{\bar{q}\sigma_{\mu\nu}G_{\mu\nu} q}}
\newcommand{\sGs}{g\braket{\bar{s}\sigma_{\mu\nu}G_{\mu\nu} s}}

\usepackage{graphicx}
\usepackage[figuresright]{rotating}

\newcommand{\AmS}{{\protect\the\textfont2
  A\kern-.1667em\lower.5ex\hbox{M}\kern-.125emS}}

\title{
The lattice QCD simulation of the quark-gluon mixed condensate $\qGq$ 
at finite temperature and the phase transition of QCD
}

\author{
T. Doi%
\address[RBRC]{RIKEN BNL Research Center, 
Brookhaven National Laboratory,
Upton, New York 11973, USA}%
\thanks{e-mail: doi@quark.phy.bnl.gov;
The Monte Carlo simulations have been performed on
NEC SX-5 at Osaka University and 
IBM Regatta at Tokyo Institute of Technology.
},
N. Ishii%
\address[titech]{
Dept. of Physics, Tokyo Institute of Technology,
Ohokayama 2-12-1, Meguro, Tokyo 152-8551, Japan}, 
M. Oka%
\addressmark[titech]
and
H. Suganuma%
\addressmark[titech]
} 

\begin{document}

\begin{abstract}

The thermal effects on the quark-gluon 
mixed condensate $\qGq$, which is another chiral order parameter,
are studied using the SU(3)$_c$ lattice QCD 
with the Kogut-Susskind fermion at the quenched level.
We perform the accurate measurement of $\qGq$ as well as $\qq$ 
for  $0 \simleq T \simleq 500{\rm MeV}$.
%
%
We observe 
the sharp decrease of both the condensates 
around $T_c \simeq 280 {\rm MeV}$, 
while the thermal effects below $T_c$ are found to be weak.
We also find that the ratio 
$m_0^2 \equiv \qGq / \qq$ 
is almost independent of the temperature
even in the very vicinity of $T_c$, which
indicates that
the two condensates have 
nontrivial similarity in the chiral behaviors.
We also present the correlation between the
condensates and the Polyakov loop
to understand the vacuum structure of QCD.

\vspace*{-4mm}
\end{abstract}

\maketitle


In order to understand the nature of QCD
and the relation to hadron phenomenology, 
the nonperturbative nature
is one of the most important characteristics of QCD.
The nonperturbative phenomena, 
such as spontaneous chiral-symmetry breaking 
and color confinement,
also bring the rich structure
to the QCD phase diagram.
For example,
at high temperature, QCD is believed to exhibit
phase transition into 
QGP,
where chiral symmetry is restored and the color is deconfined.
To realize these phenomena, 
the RHIC experiments are in progress,
which attempt to produce QGP
in the laboratory.
%
%
%
For the theoretical study of the finite temperature QCD
and the phase transition of QCD,
we focus on
the thermal effects on the condensates.
In fact, condensates 
directly characterize the nontrivial 
QCD vacuum and thus can 
indicate the change of the vacuum 
structure at finite temperature.

Among various condensates,
we study
the quark-gluon mixed condensate $\qGq$ at finite temperature 
as a relevant physical quantity 
on the chiral structure of the QCD vacuum. 
Here, the mixed condensate 
is another chiral order parameter of the QCD vacuum, 
since the chirality of the quark in $\qGq$ flips as
%
%
$
\qGq
=
  g\braket{\bar{q}_R (\sigma_{\mu\nu}G_{\mu\nu}) q_L}
+ g\braket{\bar{q}_L (\sigma_{\mu\nu}G_{\mu\nu}) q_R}.
$
%
%
%
%
%
We emphasize that $\qGq$
characterizes different aspect of the QCD vacuum
from 
$\qq$, because 
$\qGq$ reflects
the direct correlation between 
color-octet components of $q$-$\bar{q}$ pairs
and 
the spontaneously generated
gluon field, 
while $\qq$ reflects only the color-singlet $q$-$\bar{q}$ components.

We further note that 
the mixed condensate is important quantity in hadron phenomenology
through the QCD sum rule framework.
In fact, it is known that 
$\qGq$ has large effects 
in the QCD sum rule 
especially for baryons, 
such as $N$-$\Delta$ splitting~\cite{Dosch} and
parity splitting~\cite{Jido:parity}.
Recently, it is also shown~\cite{penta1}  that
$\sGs / \sbs$ 
is a key quantity for
the prediction on the parity of the 
penta-quark baryon, $\Theta^+(1540)$~\cite{Nakano}.

To study the thermal effects on $\qGq$,
we use lattice QCD Monte Carlo simulation, which 
is the direct and nonperturbative calculation from QCD.
So far, the lattice QCD studies for $\qGq$ have been
performed at zero temperature 
by a pioneering but rather preliminary 
work~\cite{K&S}, and 
by the recent works
of our group~\cite{DOIS:qGq} using the KS fermion,
and of another group~\cite{twc:qGq} using the Domain-Wall fermion.
At finite temperature, however, there has been no 
result on $\qGq$ except for 
our early reports~\cite{DOIS:T}.
Therefore, we present the extensive results
of the thermal effects on $\qGq$
with the analysis near the critical temperature~\cite{DOIS:T2}.


We evaluate the condensates 
$\qGq$ as well as $\qq$
using the SU(3)$_c$ lattice QCD at the quenched level.
The Monte Carlo simulations are performed with the 
standard Wilson action for $\beta=6.0, 6.1$ and $6.2$.
The lattice units are obtained as
$a^{-1} =$ $1.9,$ $2.3$ and $2.7 {\rm GeV}$ for 
$\beta =$ $6.0,$ $6.1$ and $6.2$, respectively, 
so as to reproduce the string tension 
$\sqrt{\sigma} = 427 {\rm MeV}$~\cite{DOIS:T2}.
We perform the calculation
at various temperatures as $0 \simleq T \simleq 500{\rm MeV}$ 
using the following lattices, 

\noindent
\begin{tabular}{cl}
(i)   & $\beta = 6.0$,\  $16^3\times N_t\ (N_t=16,12,10,8,6,4)$,\\
(ii)  & $\beta = 6.1$,\  $20^3\times N_t\ (N_t=20,12,10,8,6)$, \\
(iii) & $\beta = 6.2$,\  $24^3\times N_t\ (N_t=24,16,12,10,8)$.
\end{tabular}
%
%
%
%
%
We generate 100 gauge configurations for each lattice.
Moreover, we generate 1000 gauge configurations
in the vicinity of the phase transition point, 
namely, $20^3\times 8$ at $\beta=6.1$ and $24^3\times 10$ at
$\beta=6.2$, because the fluctuations of the condensates 
get larger near $T_c$.
For the lattices at $T > T_c$, we only use
the gauge configurations which are continuously connected
to the trivial vacuum $U_\mu=1$.
For the fermion action, we employ
the KS fermion 
to
preserve the explicit chiral symmetry
for the quark mass $m=0$, 
which is desirable
for the study of chiral order parameters.
%

In the calculation of the condensates on the lattice, we use the 
formula  as
%
%
%
%
$
a^3 \qq
= - \frac{1}{4}\sum_f {\rm Tr}\left[ \braket{q^f(x) \bar{q}^f(x)} \right]
$,
%
%
%
%
%
%
$a^5 \qGq
= - \frac{1}{4}\sum_{f,\ \mu,\nu}{\rm Tr}
        \left[ \braket{q^f(x) \bar{q}^f(x)} \sigma_{\mu\nu} G^{\rm lat}_{\mu\nu}(x)
\right],
$
%
%
%
%
where SU(4)$_f$ quark-spinor fields, $q$ and $\bar{q}$, are converted into 
spinless Grassmann KS fields 
and the gauge-link variable in the actual calculations.
The more detailed formula are given in Ref.~\cite{DOIS:qGq}.

In each configuration, we measure the condensates 
on 16 ($\beta=6.0$) or 2 ($\beta=6.1,6.2$) 
different 
points which are taken so as to be
equally spaced on the 4-dimensional lattice~\cite{DOIS:qGq,DOIS:T2}.
Therefore, 
we achieve high statistics as
1600 data at $\beta=6.0$ and 200 data at $\beta=6.1$ and $6.2$.
Note that for the lattices of 
$20^3\times 8$ ($\beta =6.1$)
and $24^3\times 10$ ($\beta =6.2$), 
we obtain 2000 data, which guarantees reliability of the results
even in the vicinity of $T_c$.

We calculate the condensates at each quark mass 
of $m \simeq 20, 35, 50$ MeV.
At each temperature,
we observe that both the condensates show 
a clear linear behavior against 
$m$, 
and therefore we fit the data with 
a linear function and determine 
the condensates in the chiral limit.
We estimate the statistical error to be 
5-7\% level using jackknife error method.
The finite-volume artifact is estimated to be
about 1\% level, through a check of 
boundary condition effects~\cite{DOIS:qGq,DOIS:T2}.


\begin{figure}[t]
\centering
\includegraphics[scale=0.18]
{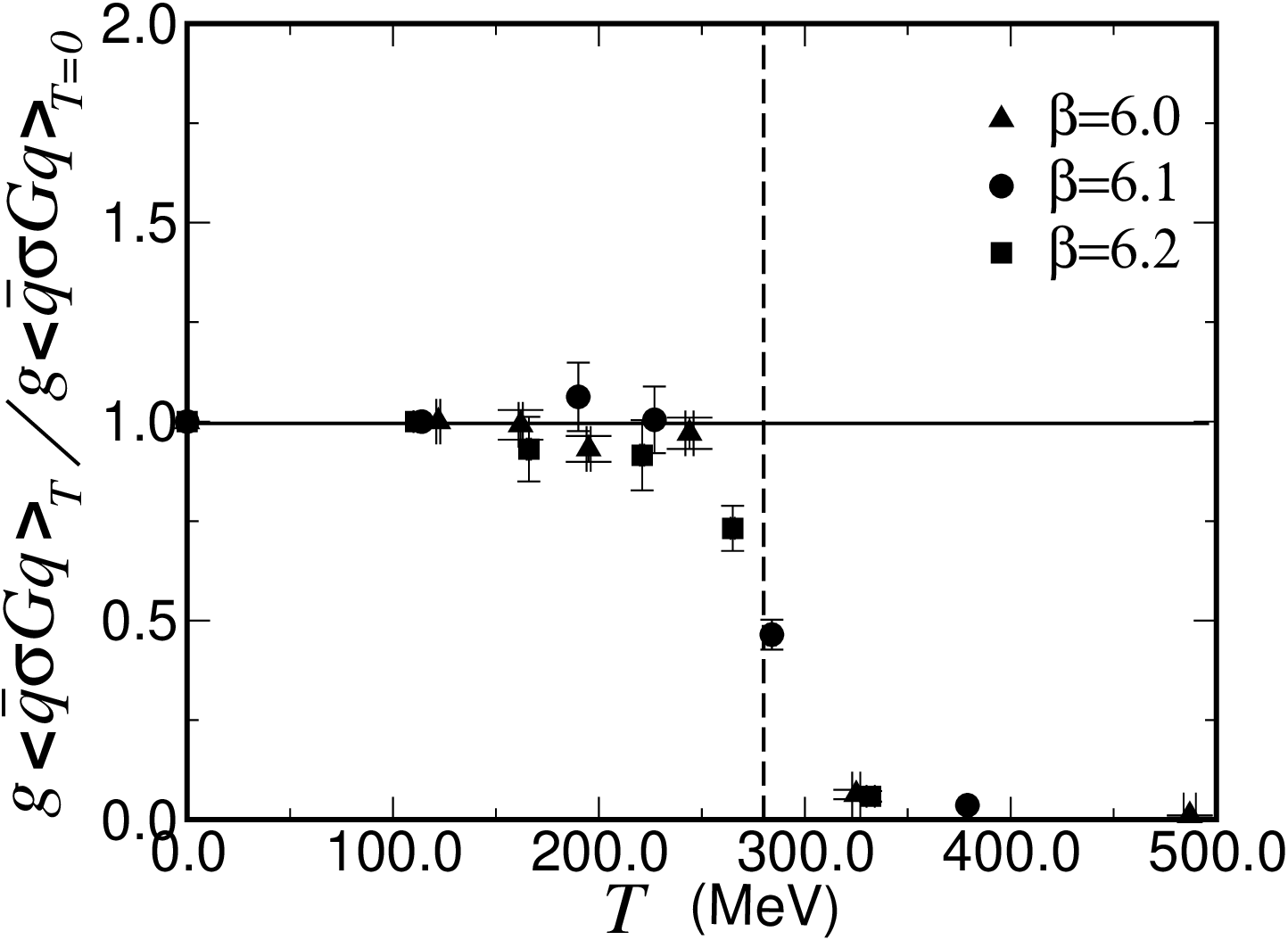}
\vspace*{-10mm}
\caption{
$\qGq_T$ normalized by $\qGq_{T=0}$
plotted against temperature $T$.
The vertical dashed line denotes the critical 
temperature $T_c \simeq 280 {\rm MeV}$
in quenched QCD.
}
\label{fig:qGq_finite_T}
\vspace*{1mm}
\includegraphics[scale=0.18]
{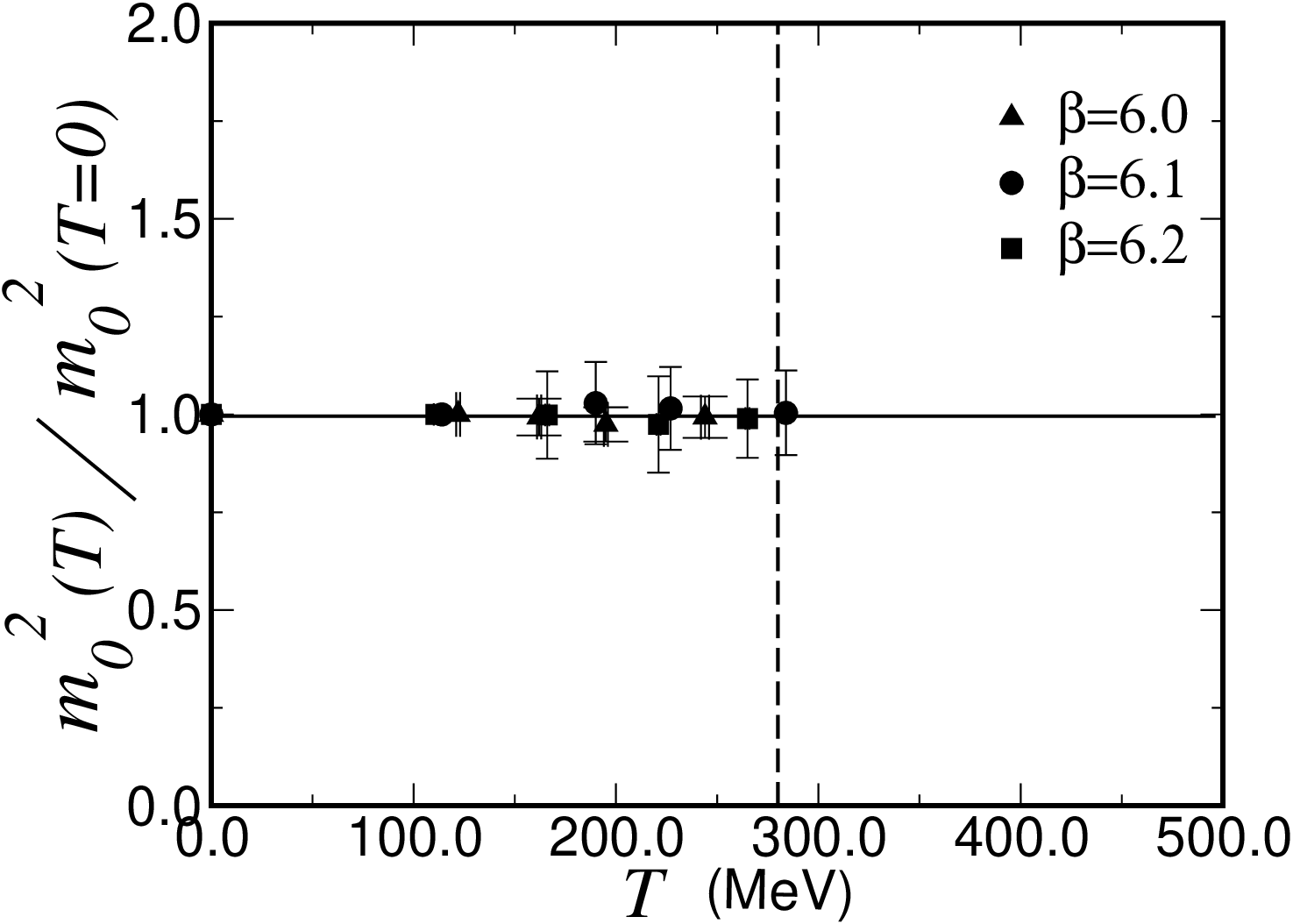}
\vspace*{-10mm}
\caption{
$m_0^2(T) \equiv \qGq_T/\qq_T$ 
normalized by $m_0^2(T=0)$ 
plotted against $T$.
This result indicates 
the same chiral behavior 
between $\qGq_T$ and $\qq_T$.
}
\label{fig:M0_finite_T}
\vspace*{-5mm}
\end{figure}

We evaluate the thermal effects
on each condensate, $\qGq$ or $\qq$, 
by taking the ratio between the values at finite 
and zero temperatures.
Note that the renormalization constants 
cancel in these ratios.
%
%
%
%
%
%
%
%
%
%
%
In figure~\ref{fig:qGq_finite_T}, we plot the thermal effects on 
$\qGq$.
We find a drastic change of $\qGq$ around the critical temperature
$T_c \simeq 280 {\rm MeV}$. 
We obtain $T_c/\sqrt{\sigma} = 0.64(4)$, which 
is consistent with the coincidence 
of the confinement/deconfinement phase transition 
and chiral-symmetry restoration.
We also find that 
the thermal effects on $\qGq$ are remarkably weak
for $T \simleq 0.9\, T_c$.
%
%
The same nontrivial features are also found for $\qq$.


We then quantitatively compare 
the thermal effects of $\qGq$ and $\qq$
by plotting 
the ratio $m_0^2(T) \equiv \qGq_T / \qq_T$
against $T$.
%
In figure~\ref{fig:M0_finite_T},
we observe that $m_0^2(T)$ is almost 
independent of the temperature, even in the very vicinity of $T_c$.
This nontrivial result
can be interpreted
that $\qGq_T$ and $\qq_T$ have
similarity in the chiral behaviors.

\begin{figure}[t]
\centering
\includegraphics[scale=0.18]
{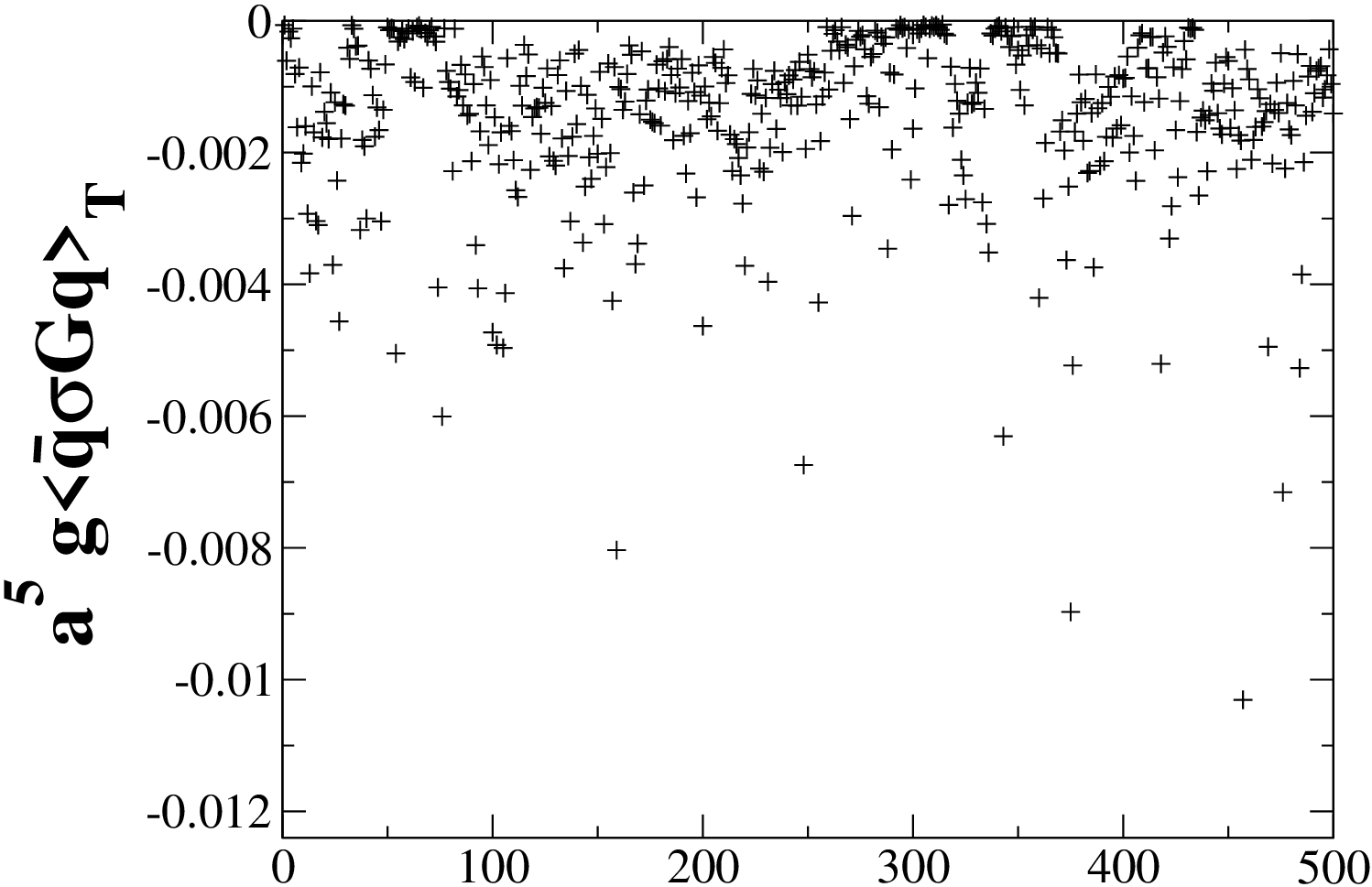}

\vspace*{5mm}
\hspace*{2mm}
\includegraphics[scale=0.18]
{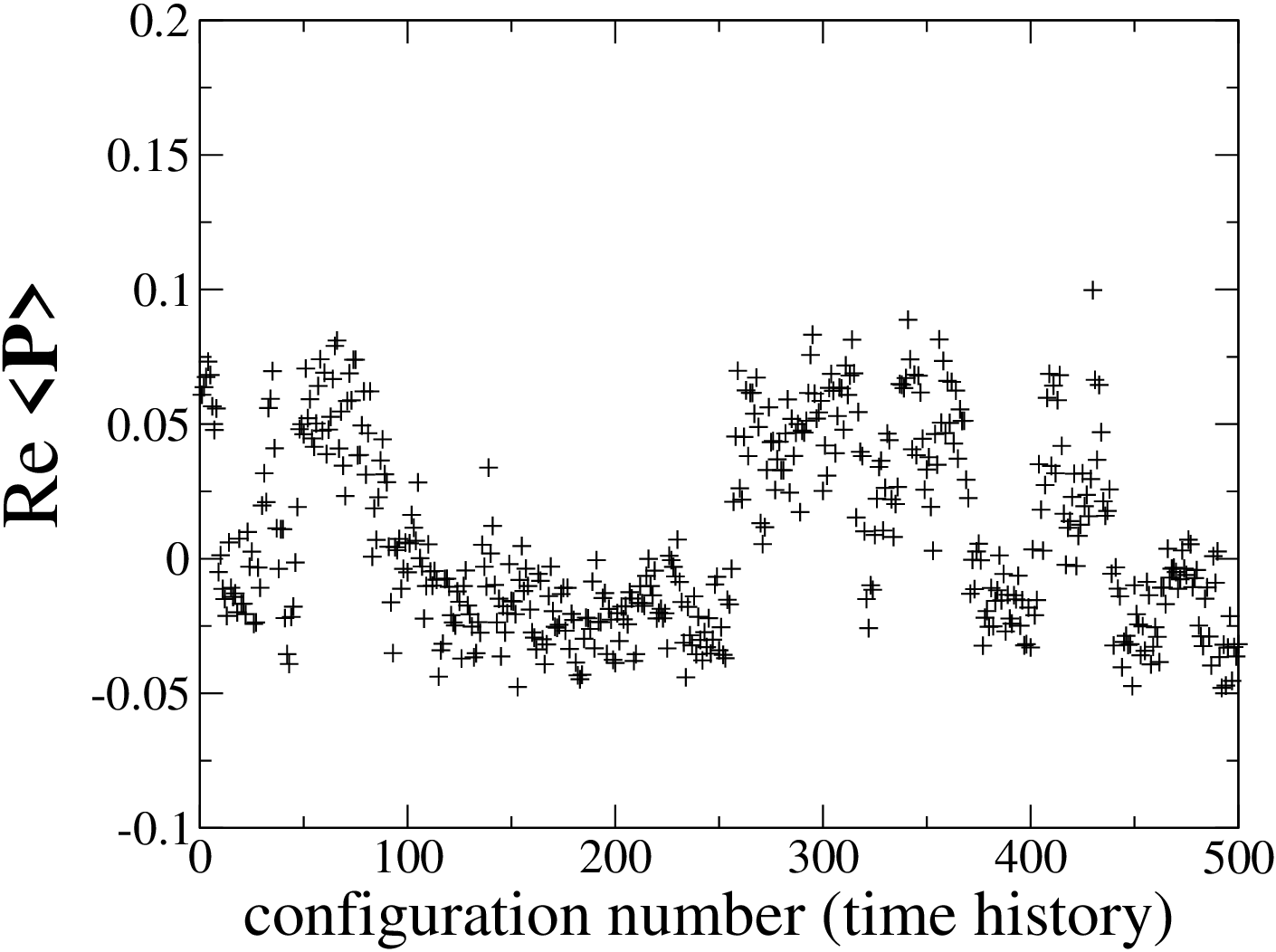}
\vspace*{-9mm}
\caption{
Time history of $\qGq_T$ (upper) and
the Polyakov loop (lower) for the  
500 configurations on $24^3\times 10$
$(\beta=6.2)$ lattice near $T_c$.
This result indicates 
strong correlation between
these two quantities.
}
\label{fig:time-history}
\vspace*{-5mm}
\end{figure}


To understand this nontrivial similarity,
we express the condensates as
$\qq =$
$\frac{1}{V}\int d\lambda' \frac{m \rho(\lambda')}{\lambda'^2 + m^2}$,
$\qGq =
\frac{1}{V}\int d\lambda' \frac{m \rho(\lambda')}{\lambda'^2 + m^2}
\braket{\lambda'|\sigma_{\mu\nu}G_{\mu\nu}|\lambda'},$
where $|\lambda\ket$ denotes the eigenvector of the Dirac operator
and 
$\rho(\lambda)$ the spectral density on $\lambda$~\cite{Banks-Hands}.
Using these formula, the common thermal behavior is understood 
in the way
that 
the thermal effects are dominated by
$\rho(\lambda)|_{\lambda=0}$ and 
$\braket{\lambda|\sigma_{\mu\nu}G_{\mu\nu}|\lambda}|_{\lambda=0}$
has remarkably weak dependence on $T$.

This result further implies that the 
thermal effects hardly appear 
in the local gluon field strength,
but appear in the global structure of 
the vacuum 
such as topological quantities. 
To proceed, we plot
the time history of $\qGq$ and 
the Polyakov loop in figure~\ref{fig:time-history}, 
and find the strong correlation
between $\qGq$ and the Polyakov loop.
For the gauge configuration with a small (large) value of $\qGq$, 
the Polyakov loop takes a large (small) value.
One of the possible explanation is that
the thermal effects
are dominated by the quark propagation 
making a circuit in time direction,
which can directly reflect the 
topological structure of the vacuum.



In summary, 
we have studied thermal effects on
$\qGq$ 
using the SU(3)$_c$ lattice QCD with  
the KS fermion at the quenched level.
We have observed  a clear signal of chiral restoration 
as a sharp decrease of $\qGq$ as well as $\qq$, while 
the thermal effects have been found to be small
for $T \simleq 0.9\, T_c$.
We have also found that 
$m_0^2(T) \equiv \qGq_T /\qq_T$
is almost independent of $T$ in the entire 
region up to $T_c$,
which indicates that these chiral condensates show
a common thermal behavior.
%
The strong correlation between the condensates and the  Polyakov loop 
has been also observed.
For further studies,
full QCD lattice calculations are 
interesting to analyze 
dynamical quark effects on the condensates.





\vspace*{-2mm}

\begin{thebibliography}{9}

\vspace*{-1mm}


\bibitem{Dosch}     
{H.G.~Dosch, M.~Jamin, and S.~Narison,
 Phys. Lett. {\bf B220} (1989) 251;
W-Y.P.~Hwang and K.-C.~Yang,
 Phys. Rev. {\bf D49} (1994) 460. }

\bibitem{Jido:parity} {D.~Jido, N.~Kodama and M.~Oka, 
                        Phys. Rev. {\bf D54} (1996) 4532;
                        D. Jido and M. Oka, hep-ph/9611322.}

\bibitem{penta1}     {J.~Sugiyama, T.~Doi and M.~Oka,
                        Phys. Lett. {\bf B581} (2004) 167.}

\bibitem{Nakano}     {T.~Nakano et al., Phys. Rev. Lett. {\bf 91} (2003) 012002.}


\bibitem{K&S}       {M.~Kremer and G.~Schierholz,
                        Phys. Lett. {\bf B194} (1987) 283 .}

\bibitem{DOIS:qGq}  {T.~Doi, N.~Ishii, M.~Oka and H.~Suganuma,
                        Phys. Rev. {\bf D67} (2003) 054504;
                        Proc. of   
                        ``Quark Confinement and the Hadron Spectrum V'',
                        (World Scientific, 2003) p.381}

\bibitem{twc:qGq}   {T.W.~Chiu and T.H.~Hsieh,
                        Nucl. Phys. {\bf B673} (2003) 217.}

\bibitem{DOIS:T}  {T.~Doi, N.~Ishii, M.~Oka, H.~Suganuma,
                        Nucl. Phys. {\bf A721} (2003) 934;
                        Prog. Theor. Phys. Suppl. {\bf 151} (2003) 161;
                        Nucl. Phys. {\bf B129} (Proc. Suppl.) (2004) 566;
%
                        Proc. of 
                        ``Color Confinement and Hadrons in Quantum Chromodynamics'',
                        (World Scientific, 2004) p.398.}

\bibitem{DOIS:T2}  {T.~Doi, N.~Ishii, M.~Oka and H.~Suganuma,
                        Phys. Rev. {\bf D70} (2004) 034510.}



\bibitem{Banks-Hands}       {T.~Banks and A.~Casher,
                        Nucl. Phys. {\bf B169} (1980) 103;
			S.~Hands, J.B.~Kogut and A.~Kocic,
                        Nucl. Phys. {\bf B357} (1991) 467.}

\end{thebibliography}
\end{document}